\documentclass[twocolumn,floatfix,superscriptaddress,showpacs]{revtex4}
\usepackage{graphicx}
\usepackage{amsmath}
\usepackage{bm}
\begin{document}

\title{Demonstration of one-parameter scaling at the Dirac point in graphene}
\author{J. H. Bardarson}
\affiliation{Instituut-Lorentz, Universiteit Leiden, P.O. Box 9506, 2300 RA Leiden, The Netherlands}
\author{J. Tworzyd{\l}o}
\affiliation{Institute of Theoretical Physics, Warsaw University, Ho\.{z}a 69, 00--681 Warsaw, Poland}
\author{P. W. Brouwer}
\affiliation{Physics Department, Arnold Sommerfeld Center for Theoretical Physics, Ludwig-Maximilans-Universit\"{a}t, 80333 Munich, Germany}
\affiliation{Laboratory of Atomic and Solid State Physics, Cornell University, Ithaca 14853, USA}
\author{C. W. J. Beenakker}
\affiliation{Instituut-Lorentz, Universiteit Leiden, P.O. Box 9506, 2300 RA Leiden, The Netherlands}
\date{May 2006}
\begin{abstract}
We numerically calculate the conductivity $\sigma$ of an undoped graphene sheet (size $L$) in the limit of vanishingly small lattice constant. We demonstrate one-parameter scaling for random impurity scattering and determine the scaling function $\beta(\sigma)=d\ln\sigma/d\ln L$.  Contrary to a recent prediction, the scaling flow has no fixed point ($\beta>0$) for conductivities up to and beyond the symplectic metal-insulator transition. Instead, the data supports an alternative scaling flow for which the conductivity at the Dirac point increases logarithmically with sample size in the absence of intervalley scattering --- without reaching a scale-invariant limit.  
\end{abstract}
\maketitle

Graphene provides a new regime for two-dimensional quantum transport \cite{Cas06,Gei07,Kat07}, governed by the absence of backscattering of Dirac fermions  \cite{And98}. A counterintuitive consequence is that adding disorder to a sheet of undoped graphene initially {\em increases\/} its conductivity \cite{Tit06,Ryc06}. Intervalley scattering at stronger disorder strengths enables backscattering \cite{McC06}, eventually leading to localization and to a vanishing conductivity in the thermodynamic limit \cite{Ale06,Alt06}. Intervalley scattering becomes less and less important if the disorder is more and more smooth on the scale of the lattice constant $a$. The fundamental question of the new quantum transport regime is how the conductivity $\sigma$ scales with increasing system size $L$ if intervalley scattering is suppressed.

\begin{figure}[tb]
\includegraphics[width=0.7\columnwidth]{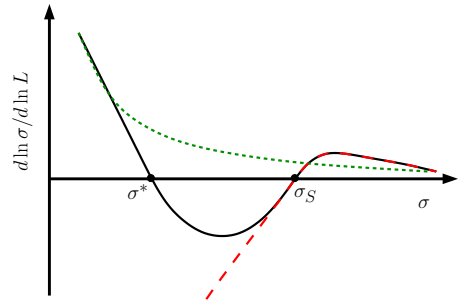} 
\caption{Two scenarios for the scaling of the conductivity $\sigma$ with sample size $L$ at the Dirac point in the absence of intervalley scattering. The black solid curve with two fixed points is proposed in Ref.\ \cite{Ost07}, the green dotted curve without a fixed point is an alternative scaling supported by the numerical data presented in this work. For comparison, we include as a red dashed curve the scaling flow in the symplectic symmetry class, which has a metal-insulator transition at $\sigma_{S}\approx 1.4$ \cite{Mar06}.}
\label{fig_beta}
\end{figure}

In usual disordered electronic systems, the hypothesis of {\em one-parameter scaling\/} plays a central role in our conceptual understanding of the metal-insulator transition \cite{Lee85,Efe97}. According to this hypothesis, the logarithmic derivative $d\ln \sigma/d\ln L=\beta(\sigma)$ is a function only of $\sigma$ itself \cite{note1} --- irrespective of the sample size or degree of disorder. A positive $\beta$-function means that the system scales towards a metal with increasing system size, while a negative $\beta$-function means that it scales towards an insulator. The metal-insulator transition is at $\beta=0$, $\beta'>0$. In a two-dimensional system with symplectic symmetry, such as graphene, one would expect a monotonically increasing $\beta$-function with a metal-insulator transition at \cite{Mar06} $\sigma_{S}\approx 1.4$ (see Fig.\ \ref{fig_beta}, green dashed curve).

Recent papers have argued that graphene might deviate in an interesting way from this simple expectation. Nomura and MacDonald \cite{Nom07} have emphasized that the very existence of a $\beta$-function in undoped
graphene is not obvious, in view of the diverging Fermi wave length at the Dirac point. Assuming that one-parameter scaling does hold, Ostrovsky,
Gornyi, and Mirlin \cite{Ost07} have proposed the scaling flow of Fig.\ \ref{fig_beta} (black solid curve). Their $\beta$-function
implies that $\sigma$ approaches a universal, scale invariant value $\sigma^{\ast}$ in the large-$L$ limit, being the hypothetical quantum
critical point of a certain field theory. This field theory differs from the symplectic sigma model by a topological term \cite{Ost07,Ryu07}. The quantum critical point could not be derived from the weak-coupling theory of Ref.\ \cite{Ost07}, but its existence was rather concluded from the analogy to the effect of a topological term in the field theory of the quantum Hall effect \cite{Efe97,Lud94}. The precise value of $\sigma^{\ast}$ is therefore unknown, but it is well constrained \cite{Ost07}: From below by the ballistic limit $\sigma_{0}=1/\pi$ \cite{Kat06,Two06,note0} and from above by the unstable fixed point $\sigma_{S}\approx 1.4$.

In this work we present a numerical test firstly, of the existence of one-parameter scaling, and secondly of the scaling prediction of Ref.\ \cite{Ost07} against an alternative scaling flow, a positive $\beta$ without a fixed point (green dotted curve in Fig.\ \ref{fig_beta}). For such a test it is crucial to avoid the finite-$a$ effects of intervalley scattering that might drive the system to an insulator before it can reach the predicted scale invariant regime. We accomplish this by starting from the Dirac equation, being the $a\rightarrow 0$ limit of the tight-binding model on a honeycomb lattice. We have developed an efficient transfer operator method to solve this equation, which we describe before proceeding to the results.

The single-valley Dirac Hamiltonian reads
\begin{equation}
  \label{eq:H}
  H = v{\bf p}\cdot\bm{\sigma} + V(x) + U(x,y).
\end{equation}
The vector of Pauli matrices $\bm\sigma$ acts on the sublattice index of the spinor $\Psi$, $\bm{p}=-i\hbar\partial/\partial\bm{r}$ is the momentum operator, and $v$ is the velocity of the massless excitations. The disorder potential $U(\bm{r})$ varies randomly in the strip $0<x<L$, $0<y<W$ (with zero average, $\langle U\rangle=0$). This disordered strip is connected to highly doped ballistic leads, according to the doping profile $V(x)=0$ for $0 < x < L$, $V(x)\rightarrow-\infty$ for $x < 0$ and $x > L$. We set the Fermi energy at zero (the Dirac point), so that the disordered strip is undoped. The disorder strength is quantified by the correlator
\begin{equation}
K_{0}=\frac{1}{(\hbar v)^{2}}\int d\bm{r}'\,\langle U(\bm{r})U(\bm{r}')\rangle.\label{eq:K0}
\end{equation}

Following Refs.\ \cite{Tit06,Che06}, we work with a transfer operator representation of the Dirac equation $H\Psi=0$ at zero energy. We discretize $x$ at the $N$ points $x_{1},x_{2},\ldots x_{N}$ and represent the impurity potential by $U(\bm{r})=\sum_{n}U_{n}(y)\delta(x-x_{n})$. Upon multiplication by $i\sigma_{x}$ the Dirac equation in the interval $0<x<L$ takes the form
\begin{equation}
\hbar v\frac{\partial}{\partial x}\Psi_{x}(y)=\bigl[vp_{y}\sigma_{z}-i\sigma_{x}\sum_{n}U_{n}(y)\delta(x-x_{n})\bigr]\Psi_{x}(y).\label{Diractransfer}
\end{equation}
The transfer operator ${\cal M}$, defined by $\Psi_{L}={\cal M}\Psi_{0}$, is given by the operator product
\begin{align}
{\cal M} ={} & {\cal P}_{L,x_{N}}{\cal K}_{N}{\cal P}_{x_{N},x_{N-1}}\cdots {\cal K}_{2}{\cal P}_{x_{2},x_{1}}{\cal K}_{1}{\cal P}_{x_{1},0},\label{Mproduct}\\
{\cal P}_{x,x'}={} &\exp[(1/\hbar)(x-x')p_{y}\sigma_{z}],\label{Pdef}\\
{\cal K}_{n}={} &\exp[-(i/\hbar v)U_{n}\sigma_{x}].\label{Kdef}
\end{align}
The operator ${\cal P}$ gives the decay of evanescent waves between two scattering events, described by the operators ${\cal K}_{n}$. For later use we note the current conservation relation
\begin{equation}
{\cal M}^{-1}=\sigma_{x}{\cal M}^{\dagger}\sigma_{x}.\label{currentconserve}
\end{equation}

We assume periodic boundary conditions in the $y$-direction, so that we can represent the operators in the basis 
\begin{equation}
  \label{eq:modes}
  \psi_{k}^{\pm} = \frac{1}{\sqrt{W}}e^{iq_k y}|\pm\rangle,\;\; q_{k}=\frac{2\pi k}{W},\;\; k=0,\pm 1,\pm 2\ldots.
\end{equation}
The spinors $|+\rangle= 2^{-1/2}\binom{1}{1}$, $|-\rangle= 2^{-1/2}\binom{1}{-1}$ are eigenvectors of $\sigma_{x}$. In this basis, $(p_{y})_{kk'}=\hbar q_{k}\delta_{kk'}$ is a diagonal operator, while $(U_{n})_{kk'}=W^{-1}\int dy\,U_{n}(y)\exp[i(q_{k'}-q_{k})y]$ is nondiagonal. We work with finite-dimensional transfer matrices by truncating the transverse momenta $q_k$ at $|k| = M$.

The transmission and reflection matrices $\bm{t}$, $\bm{r}$ are determined as in Ref.\ \cite{Two06}, by matching the amplitudes of incoming, reflected, and transmitted modes in the heavily doped graphene leads to states in the undoped strip at $x=0$ and $x=L$. This leads to the set of linear equations
\begin{subequations}
\label{matching}
\begin{align}
& \sum_{k}\left[\delta_{kk'}\psi_{k}^{+}(y)+r_{kk'}\psi_{k}^{-}(y)\right]=\Psi_{0}(y),\label{matchinga}\\
& \sum_{k}t_{kk'}\psi_{k}^{+}(y)=\Psi_{L}(y)={\cal M}\Psi_{0}(y).\label{matchingb}
\end{align}
\end{subequations}
Using the current conservation relation \eqref{currentconserve} we can solve Eq.\ \eqref{matching} for the transmission matrix, 
\begin{equation}
\begin{pmatrix}
1-\bm{r}\\ 1+\bm{r}
\end{pmatrix}={\cal M}^{\dagger}\begin{pmatrix}\bm{t}\\ \bm{t}\end{pmatrix}\Rightarrow \bm{t}^{-1}=
\langle +|{\cal M}^{\dagger}|+\rangle.\label{tnMrelation}
\end{equation}
The transmission matrix determines the conductance $G=(4e^{2}/h)\,{\rm Tr}\,\bm{tt}^{\dagger}$, and hence the dimensionless conductivity $\sigma=(h/4e^{2})(L/W)G$. The average conductivity $\langle\sigma\rangle$ is obtained by sampling some $10^{2}-10^{3}$ realizations of the impurity potential.

Because the transfer matrix ${\cal P}$ has both exponentially small and exponentially large eigenvalues, the matrix multiplication \eqref{Mproduct} is
numerically unstable. As in Ref.\ \cite{Tam91}, we stabilize the product of transfer matrices by transforming it into a composition of unitary scattering matrices, involving only eigenvalues of unit absolute value.

We model the disorder potential $U(\bm{r})=\sum_{n=1}^{N}\gamma_n\delta(x-x_{n})\delta(y-y_{n})$ by a collection of $N$ isolated impurities distributed uniformly over a strip $0<x<L$, $0<y<W$. (An alternative model of a continuous Gaussian random potential is discussed at the end of the paper.) The strengths $\gamma_n$ of the scatterers are uniform in the interval $[-\gamma_0, \gamma_0]$. The number $N$ sets the average separation $d=(WL/N)^{1/2}$ of the scatterers. The cut-off $|k|\leq M$ imposed on the transverse momenta $q_k$ limits the spatial resolution $\xi\equiv W/(2M+1)$ of plane waves $\propto e^{iq_{k}y\pm q_{k}x}$ at the Dirac point. The resulting finite correlation lengths of the scattering potential in the $x$- and $y$-directions scale with $\xi$, but they are not determined more precisely. The disorder strength \eqref{eq:K0} evaluates to $K_{0}=\frac{1}{3}\gamma_{0}^{2}(\hbar v d)^{-2}$, independent of the correlation lengths. We scale towards an infinite system by increasing $M\propto L$ at fixed disorder strength $K_{0}$, scattering range $\xi/d$, and aspect ratio $W/L$.

\begin{figure}[tb]
\includegraphics[width=0.7\columnwidth, angle=-90]{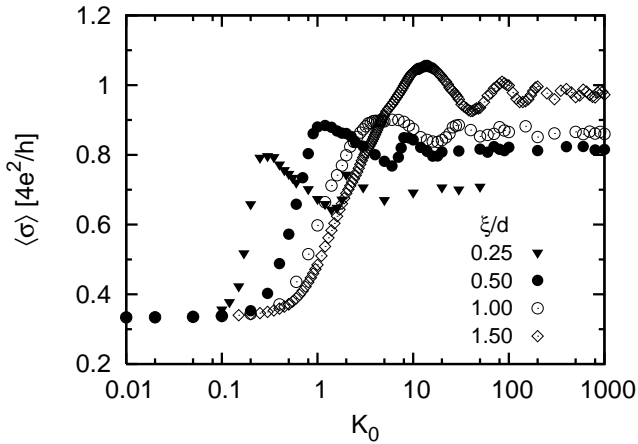} 
\caption{Disorder strength dependence of the average conductivity for a fixed system size ($W=4\,L=40\,d$) and four values of the scattering range.}
\label{fig_sigmavsK0}
\end{figure}

This completes the description of our numerical method. We now turn to the results. In Fig.\ \ref{fig_sigmavsK0} we first show the dependence of the average conductivity on $K_{0}$ for a fixed system size. As in the tight-binding model of Ref.\ \cite{Ryc06}, disorder increases the conductivity above the ballistic value. This impurity assisted tunneling \cite{Tit06} saturates in an oscillatory fashion for $K_{0}\gg 1$ (unitary limit \cite{Khv06,Ost06}). In the tight-binding model \cite{Ryc06} the initial increase of $\sigma$ was followed by a rapid decay of the conductivity for $K_{0}\gtrsim 1$, presumably due to Anderson localization. The present model avoids localization by eliminating intervalley scattering from the outset.

\begin{figure}[tb]
\includegraphics[width=0.7\columnwidth, angle=-90]{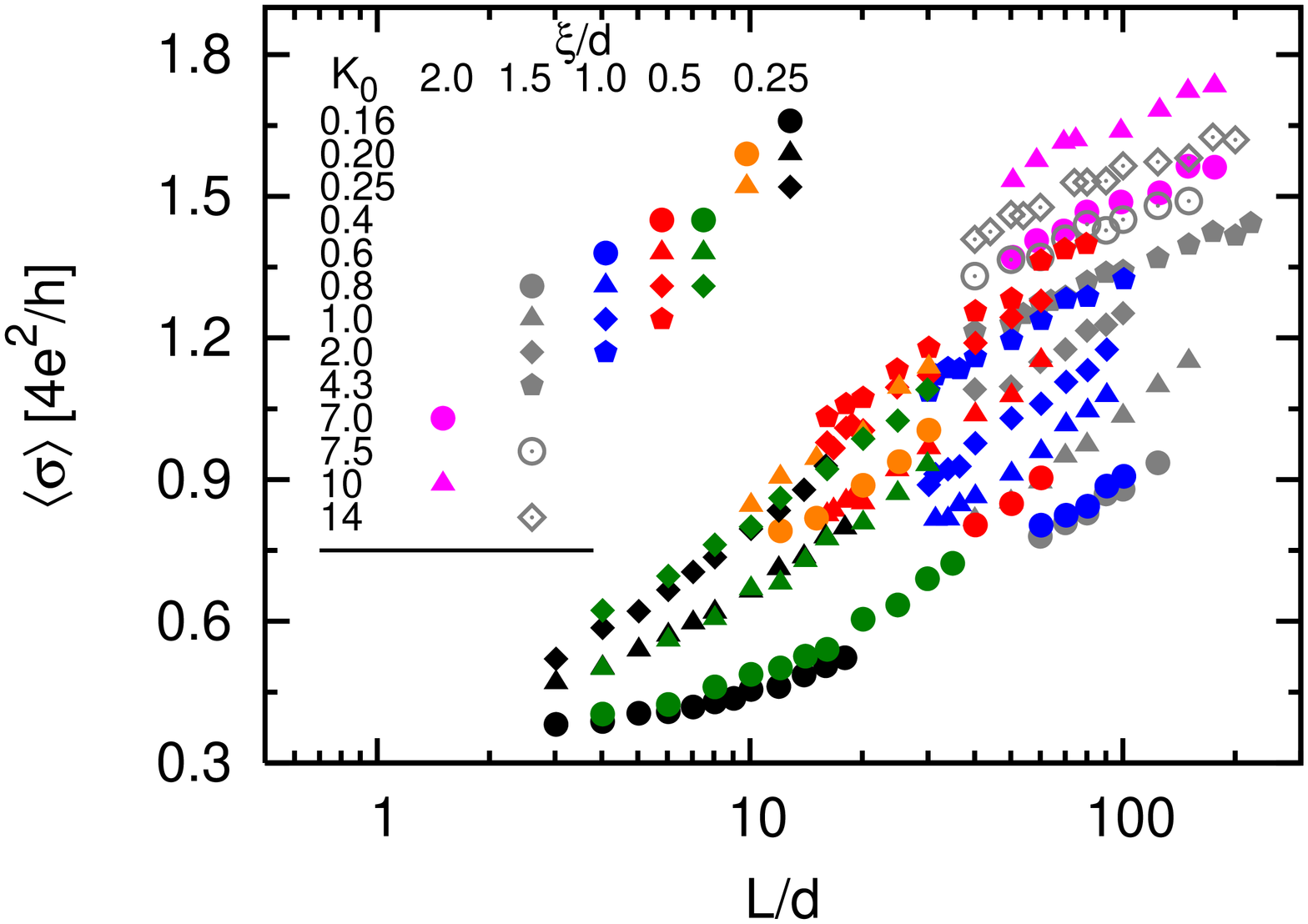}
    
\includegraphics[width=0.7\columnwidth, angle=-90]{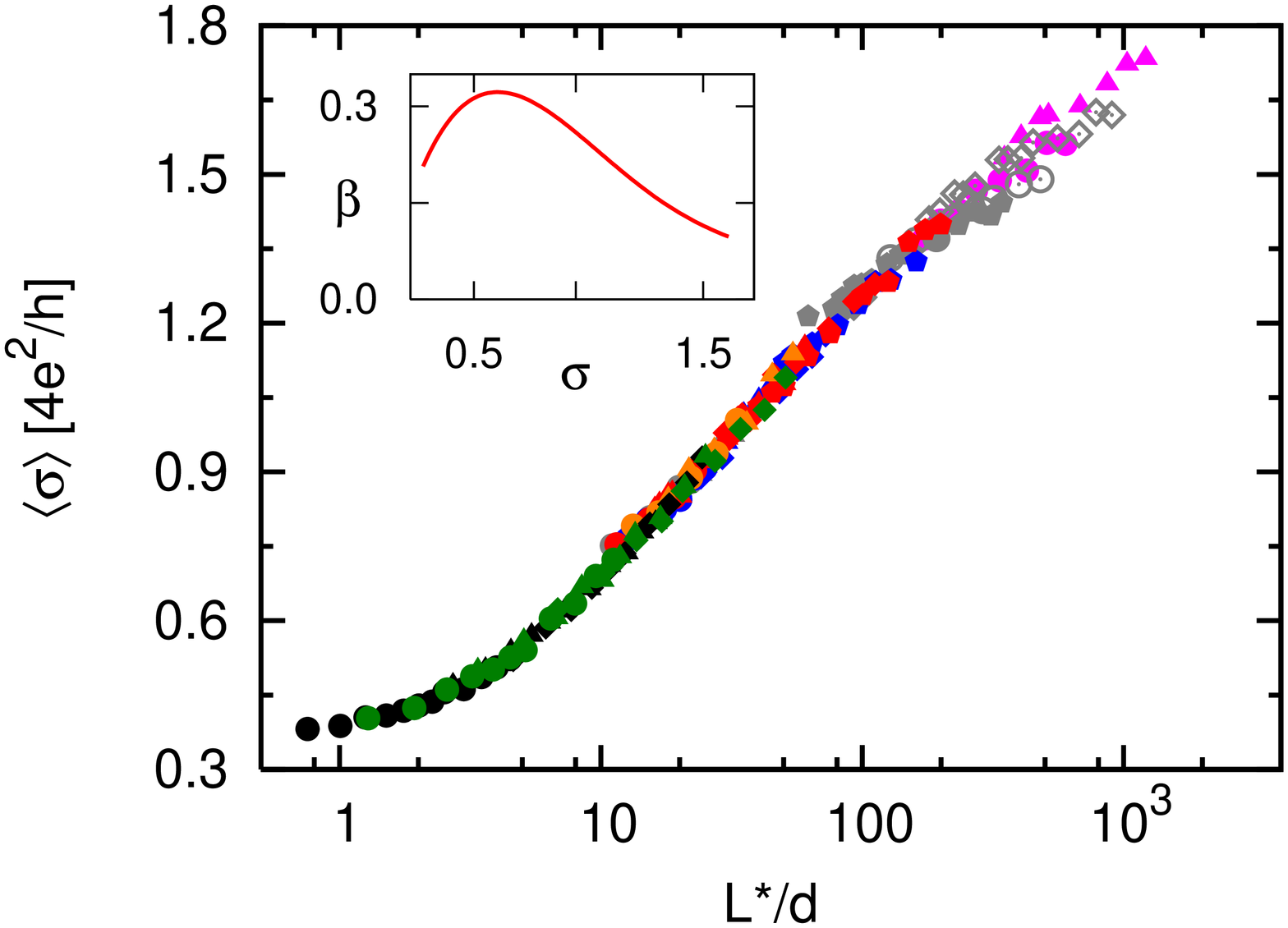} 
\caption{System size dependence of the average conductivity, for $W/L=4$ (black and green solid symbols) and $W/L=1.5$ (all other
symbols) and various combinations of $K_{0}$ and $\xi/d$. The top panel shows the raw data. In the bottom panel the data sets have been given a horizontal offset, to demonstrate the existence of one-parameter scaling. The inset shows the resulting $\beta$-function.}
\label{fig_cond}
\end{figure}

The system size dependence of the average conductivity is shown in Fig.\ \ref{fig_cond}, for various combinations of
disorder strength and scattering range. We take $W/L$ sufficiently large that we have reached an aspect-ratio
independent scaling flow and $L/d$ large enough that the momentum cut-off $M > 25$. The top panel shows the data sets as a function
of $L/d$. The increase of $\sigma$ with $L$ is approximately logarithmic, $\langle\sigma\rangle= {\rm constant}+0.25\ln L$, much slower than the $\sqrt{L}$ increase obtained in Ref.~\onlinecite{Tit06} in the absence of mode mixing.

If one-parameter scaling holds, then it should be possible to rescale the length $L^{\ast}\equiv f(K_{0},\xi/d)L$ such that the data sets collapse onto a single smooth curve when plotted as a function of $L^{\ast}/d$. (The function $f\equiv d/l^{\ast}$ determines the effective mean free path $l^{\ast}$, so that $L^{\ast}/d\equiv L/l^{\ast}$.) The bottom panel in Fig.\ \ref{fig_cond} demonstrates that, indeed, this data collapse occurs. The resulting $\beta$-function is plotted in the inset. Starting from the ballistic limit \cite{note0} at $\sigma_{0}=1/\pi$, the $\beta$-function first rises until $\sigma\approx 0.6$, and then decays to zero without becoming negative. For $\sigma> \sigma_{S}\approx 1.4$ the decay $\propto 1/\sigma$ is as expected for a diffusive system in the symplectic symmetry class. The positive $\beta$-function in the interval $(\sigma_{0},\sigma_{S})$ precludes the flow towards a scale-invariant conductivity predicted in Ref.\ \cite{Ost07}.

\begin{figure}
\includegraphics[width=0.7\columnwidth, , angle=-90]{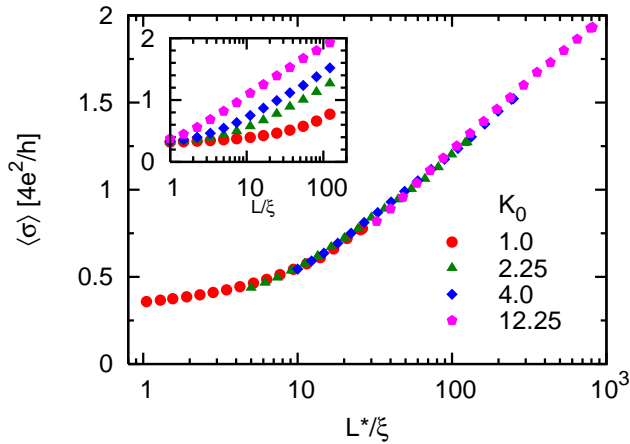} 
\caption{System size dependence of the average conductivity in the continuous potential model, for several values of $K_{0}$. The inset shows the raw data, while the data sets in the main plot have a horizontal offset to demonstrate one-parameter scaling when $L\gtrsim 5\,\xi$.
\label{fig_cont}}
\end{figure}

The model of isolated impurities considered so far is used in much of the theoretical literature, whereas experimentally a continuous random potential is more realistic \cite{Nom07}. We have therefore also performed numerical simulations for a random potential landscape with Gaussian correlations \cite{Ner95},
\begin{equation}
  \langle U(\bm{r}) U(\bm{r}') \rangle =
   K_{0}\frac{(\hbar v)^2}{2\pi \xi^2} e^{-|\bm{r} - \bm{r}'|^2/2\xi^2}.
\end{equation}
The discrete points $x_{1}, x_{2} \ldots x_N$ in the operator product (\ref{Mproduct}) are taken equidistant with spacing $\delta x = L/N$, and
\begin{equation}
  U_n(y) = \int_{x_n - \delta x/2}^{x_n + \delta x/2} dx\, U(x,y).
\end{equation}
We take $M$, $N$, and $W/L$ large enough that the resulting conductivity no longer depends on these parameters. We then scale towards
larger system sizes by increasing $L/\xi$ and $W/\xi$ at fixed $K_{0}$. No saturation of $\sigma$ with increasing $K_{0}$ is observed for
the continuous random potential (as expected, since the unitary limit is specific for isolated scatterers \cite{Khv06,Ost06}). Fig.\
\ref{fig_cont} shows the size dependence of the conductivity --- both the raw data as a function of $L$ (inset) as well as the rescaled data as a function of $L^{\ast}\equiv g(K_{0})L$. Single-parameter scaling applies for $L\gtrsim 5\,\xi$, where $\langle\sigma\rangle={\rm constant}+0.32\ln L$. The prefactor of the logarithm is about 25\% larger than in the model  of isolated impurities (Fig.\ \ref{fig_cond}), which is within the numerical uncertainty.

In conclusion, we have demonstrated that the central hypothesis of the scaling theory of quantum transport, the existence of one-parameter scaling, holds in graphene. The scaling flow which we find (green dotted curve in Fig.\ \ref{fig_beta})  is qualitatively different both from what would be expected for conventional electronic systems (red dashed curve) and also from what has been predicted \cite{Ost06} for graphene (black solid curve). Our scaling flow has no fixed point, meaning that the conductivity of undoped graphene keeps increasing with increasing disorder in the absence of intervalley scattering. The fundamental question ``what is the limiting conductivity $\sigma_{\infty}$ of an infinitely large undoped carbon monolayer'' has therefore three different answers: $\sigma_{\infty}=1/\pi$ in the absence of any disorder \cite{Kat06,Two06}, $\sigma_{\infty}=\infty$ with disorder that does not mix the valleys (this work), and $\sigma_{\infty}=0$ with intervalley scattering \cite{Ale06,Alt06}.

We thank C. Mudry and M. Titov for valuable discussions. This research was supported by the Dutch Science Foundation NWO/FOM, the European
Community's Marie Curie Research Training Network (contract MRTN-CT-2003-504574, Fundamentals of Nanoelectronics), and by the Packard
Foundation.

{\em Note added:} Since submission of this manuscript, similar conclusions have been reported by Nomura, Koshino, and Ryu \cite{Nom07b}.


\begin{thebibliography}{99}
\bibitem{note1} We define the $\beta$-function in terms of the ensemble averaged conductivity $\sigma$, measured in units of $4e^{2}/h$ (with the factor of four accounting for twofold spin and valley degeneracies). This is the appropriate definition for our system. For a more general definition of one-parameter scaling, one needs to scale a distribution function of conductances \cite{Efe97}.
\bibitem{note0} We call $\sigma_{0}$ the ballistic limit because it is reached in the absence of disorder, but we emphasize that it is a conductivity --- not a conductance. This is a unique property (called ``pseudodiffusive'') of graphene at the Dirac point, that its conductance scales $\propto 1/L$ like in a diffusive system even in the absence of disorder.
\end{thebibliography}
\end{document}